\documentclass[twocolumn,showpacs,preprintnumbers,amsmath,amssymb,prb]{revtex4}

\usepackage{graphicx}
\usepackage{color}

\begin{document}

\title{Quantum effects in graphene monolayers: Path-integral simulations}
\author{Carlos P. Herrero}
\author{Rafael Ram\'irez}
\affiliation{Instituto de Ciencia de Materiales de Madrid,
         Consejo Superior de Investigaciones Cient\'ificas (CSIC),
         Campus de Cantoblanco, 28049 Madrid, Spain }
\date{\today}

\begin{abstract}
Path-integral molecular dynamics (PIMD) simulations have been carried out 
to study the influence of quantum dynamics of carbon atoms on the
properties of a single graphene layer.
Finite-temperature properties were analyzed in the range from 
12 to 2000~K, by using the LCBOPII effective potential.
To assess the magnitude of quantum effects in structural and
thermodynamic properties of graphene, classical molecular
dynamics simulations have been also performed. 
Particular emphasis has been laid on the atomic vibrations along the 
out-of-plane direction.
Even though quantum effects are present in these vibrational modes,
we show that at any finite temperature classical-like motion dominates
over quantum delocalization, provided that the system size is large enough.
Vibrational modes display an appreciable anharmonicity, as derived 
from a comparison between kinetic and potential energy of the carbon atoms. 
Nuclear quantum effects are found to be appreciable in the interatomic 
distance and layer area at finite temperatures.
The thermal expansion coefficient resulting from PIMD simulations
vanishes in the zero-temperature limit, in agreement with 
the third law of thermodynamics. 
\end{abstract}



\maketitle

\section{Introduction}

In recent years there has been a growth of interest in carbon-based
materials, in particular in those formed by C atoms with $sp^2$
hybridization, as is the case of carbon nanotubes, fullerenes, 
and graphene, a two-dimensional (2D) crystal with exceptional 
electronic properties.\cite{ge07,fl11}
In the zero-temperature limit and in the absence of defects, the C-C bond 
on the graphene layer is well understood in terms of in-plane 
localized sp$^2$ bonds and delocalized out-of-plane $\pi$-like 
bonds.\cite{ca09b} The optimum structural arrangement is reached 
with a honeycomb lattice, resulting in a material with
the largest in-plane elastic constants known to date.
Departures from this flat structure may appreciably affect
the atomic scale properties of the graphene layer and can modify 
the properties of this material.
Although the main interest on graphene is due to its extraordinary
electronic properties,\cite{ca09b} the observation of ripples in freely
hanged samples gave rise to theoretical interest also in the
structural properties of this material.\cite{me07,fa07,ki08,am16} 
In fact, ripples or bending fluctuations have been proposed as one of the
main scattering mechanisms limiting the electronic mobility
in graphene.\cite{ka08,gu08}

There appear various reasons for a graphene sheet to depart from strict 
planarity, among them the presence of defects and
external stresses.\cite{fa07,an12}
Defects (e.g., vacancies or impurities) originate deformations
in the graphene sheet at the atomic scale, that can propagate
giving rise to long-range correlations, and
external stresses due to the boundary conditions 
cause bending and corrugation of the sheet.\cite{wa13,we13}
Even in the absence of defects and stresses, a perfect 2D crystalline 
layer in three-dimensional (3D) space cannot be stable at finite 
temperatures in the thermodynamic limit.\cite{me68}
Moreover, thermal fluctuations at finite temperatures
produce out-of-plane motion of the carbon atoms,
and even in the zero-temperature limit, quantum fluctuations associated to
zero-point motion in the out-of-plane direction will cause
a departure of strict planarity of the graphene sheet.

From a basic point of view,
understanding structural and thermal properties of 2D systems is 
a challenging problem in modern statistical physics.\cite{sa94,ne04,ta13}
It has been traditionally considered mainly in the context of biological
membranes and soft condensed matter. The great complexity of these
systems has hindered a microscopic approach based on realistic 
descriptions of the interatomic interactions.
Graphene provides us with a model system
where an atomistic description is feasible, thus opening the way to a
better comprehension of the physical properties of this kind of systems.
Thus, there has been recently a rise of interest on thermodynamic 
properties of graphene, both theoretically and 
experimentally.\cite{po12b,fo13,am14,wa16}

Finite-temperature properties of graphene have been studied 
by molecular dynamics and Monte Carlo simulations using
{\em ab-initio},\cite{sh08,an12b,ch14} 
tight binding,\cite{ca09c,he09a,ak12,le13} 
and empirical interatomic potentials.\cite{fa07,lo09,sh13,ma14,lo16,ra16}
In most applications of these methods, atomic nuclei were described as
classical particles.
To take into account the quantum character of the nuclei,
path-integral (both, molecular dynamics and Monte Carlo) simulations 
turn out to be particularly suitable.  In these methods all nuclear 
degrees of freedom can be quantized in an efficient way, permitting us 
to include quantum and thermal fluctuations in many-body systems
at finite temperatures.  This allows to carry out quantitative studies 
of anharmonic effects in condensed matter.\cite{gi88,ce95}
Recently, Brito {\em et al.}\cite{br15} have carried out path-integral
Monte Carlo simulations of a single graphene layer. These authors examined
several equilibrium properties of graphene at finite temperatures
using a supercell including 200 carbon atoms. This computational
technique has been also applied to study structural properties of a
boron nitride monolayer.\cite{ca16}
Moreover, nuclear quantum effects in graphene have been studied earlier 
by using a combination of density-functional theory and a 
quasi-harmonic approximation for the vibrational modes.\cite{mo05,sh12} 

In this paper, the path-integral molecular dynamics (PIMD) method is 
used to investigate the influence of nuclear quantum dynamics on 
structural and thermodynamic properties of graphene at temperatures
up to 2000~K. We consider simulation cells of different sizes, since
finite-size effects are expected to be very important for some equilibrium
variables, in particular the projected area of the graphene layer on 
the plane defined by the simulation box, and the atomic delocalization
(classical and quantum) in the out-of-plane direction.\cite{ga14,lo16}
Low-temperature values of these quantities are analyzed in terms of 
the third law of thermodynamics, which has to be fulfilled by the
results of the quantum simulations as $T \to 0$.
We also analyze the anharmonicity of the vibrational modes by comparing 
results for the kinetic and potential energy derived from the PIMD simulations.

Path-integral methods analogous to that employed in this work 
have been applied earlier to study nuclear
quantum effects in pure and hydrogen-doped carbon-based materials, 
as diamond\cite{he00c,ra06,he07} and graphite.\cite{he10}
Helium adsorption\cite{kw12} and diffusion of H on graphene\cite{he09a}
have been also studied earlier by using this kind of techniques.

 The paper is organized as follows. In Sec.\,II, we describe the
computational method employed in the calculations. 
Our results are presented in Sec.\,III, dealing with the interatomic 
distance, layer area, vibrational energy, and atomic delocalization.
In Sec.\,IV we discuss the thermodynamic consistency of our data
at low temperature, and in Sec.\,V we summarize the main results.

\section{Computational Method}

We employ the PIMD method to obtain equilibrium properties of graphene
at various temperatures.
This method is based on the path-integral formulation of statistical
mechanics, which is a powerful nonperturbative approach to study many-body
quantum systems at finite temperatures.
It exploits the fact that the partition function of a quantum system 
can be written in a way formally equivalent to that of a classical one, 
obtained by substituting each
quantum particle by a ring polymer consisting of $L$ (Trotter number)
classical particles, connected by harmonic springs.\cite{gi88,ce95}
Thus, the actual implementation of this procedure is based on an
isomorphism between the real quantum system and a fictitious classical 
system of ring polymers, in which each quantum particle is described by 
a polymer (corresponding to a cyclic quantum path) composed of  $L$ 
``beads''.\cite{fe72,kl90} This becomes exact in the limit $L \to \infty$. 
Details on this simulation technique can be found
elsewhere.\cite{ch81,gi88,ce95,he14}

Here we use the molecular dynamics method to sample the configuration
space of the classical isomorph of our quantum system ($N$ carbon atoms).
This is the so-called PIMD method.  We note that the dynamics in this 
computational technique is artificial, in the sense that it
does not correspond to the actual quantum dynamics of the real particles
under consideration. It is, however, useful for effectively sampling the
many-body configuration space, giving precise results for 
time-independent equilibrium properties of the quantum system.

An important question in the PIMD method is an adequate description of the
interatomic interactions, which should be as realistic as possible.
Since employing an {\em ab-initio} method would enormously restrict
the size of our simulation cell, we derive the Born-Oppenheimer 
surface for the nuclear dynamics from an effective empirical potential,
developed for carbon-based systems, namely the so-called 
LCBOPII.\cite{lo03b,lo05,gh08}
This is a long-range carbon bond order potential, which has been employed
earlier to carry out classical simulations of diamond,\cite{lo05} 
graphite,\cite{lo05}, liquid carbon,\cite{gh05} 
as well as graphene layers.\cite{fa07,za09,za10b,za11,lo16}
In particular, it was used to predict the carbon phase diagram comprising
graphite, diamond, and the liquid, showing its accuracy by comparison of the
predicted graphite-diamond line with experimental data.\cite{gh05b}
In the case of graphene, this effective potential has been found to 
give a good description of elastic properties such as the Young's
modulus.\cite{za09,po12} 
Also, the interatomic potential employed here yields at 300 K a bending 
modulus $\kappa$ = 1.6 eV for graphene at 300 K.\cite{ra16}
This value is close to the best fit to experimental and theoretical results
obtained for $\kappa$ by Lambin.\cite{la14} 
To our knowledge, this is the first time that the LCBOPII potential
is used to perform path-integral simulations of this kind of systems.

Calculations were carried out in the isothermal-isobaric ensemble,
where we fix the number of carbon atoms ($N$), the applied stress 
($P = 0$), and the temperature ($T$).
We used effective algorithms for performing PIMD simulations
in this statistical ensemble, as those described in the
literature.\cite{tu92,tu98,ma99,tu02}  In particular,
we employed staging variables to define the bead coordinates, and
the constant-temperature ensemble was generated by coupling chains
of four Nos\'e-Hoover thermostats with mass $M = \beta \hbar^2 / 5 L$ 
to each staging variable ($\beta = 1 / k_B T$).  
An additional chain of four barostats was coupled to the area 
of the simulation box to yield the required constant pressure 
(here $P = 0$).\cite{tu98,he14}

The equations of motion were integrated by employing
the reversible reference system propagator algorithm (RESPA), which allows
us to define different time steps for the integration of the fast and slow
degrees of freedom.\cite{ma96}
The time step $\Delta t$ associated to the interatomic forces was taken in 
the range between 0.5 and 1 fs, which was found to be adequate for the 
interatomic interactions, atomic masses, and temperatures
considered here, and provided appropriate convergence for the 
studied magnitudes.
For the evolution of the fast dynamical variables, including the
thermostats and harmonic bead interactions, we used a
time step $\delta t = \Delta t/4$, as in earlier PIMD
simulations.\cite{he06,he11}  
The kinetic energy $E_k$ has been calculated by using the so-called virial
estimator, which is known to have a statistical uncertainty
smaller than the potential energy of the system.\cite{he82,tu98}

Sampling of the configuration space has been carried out at temperatures
between 12~K and 2000~K.
For comparison with results of PIMD simulations, some classical
molecular dynamics (MD) simulations of graphene have been also carried out.
This is realized in our context by setting the Trotter number $L$ = 1.
For the quantum simulations,
the Trotter number $L$ was taken proportional to the inverse temperature
($L \propto 1/T$), so that $L \, T$ = 6000~K, which turns out to roughly 
keep a constant precision in the PIMD results at different 
temperatures.\cite{he06,he11,ra12}

We have considered rectangular simulation cells with similar side length
in the $x$ and $y$ directions of the $(x, y)$ reference plane, and
periodic boundary conditions were assumed.
We checked that using isotropic changes of the cell dimensions in the $NPT$ 
simulations yielded for the studied variables the same results as allowing
flexible cells (independent changes of $x$ and $y$ axis length along with
deformations of the rectangular shape).
Moreover, to check the possible influence of the boundary conditions,
we also used supercells of the primitive hexagonal cell, and 
found no change in our results.   
Cells of size up to 33600 atoms were considered for simulations at 
$T \geq$ 300~K, but at lower temperatures, smaller sizes were dealt 
with due to the fast increase in $L$. 
(Note that for $N$ = 33600 carbon atoms at 100 K, we have to handle a 
total of $N L \sim 2 \times 10^6$ classical-like beads.)  
For a given temperature, a typical simulation run consisted of
$3 \times 10^5$ PIMD steps for system equilibration, followed by
$4 \times 10^6$ steps for the calculation of ensemble average properties.
We have checked that this simulation length is much larger than the
the autocorrelation time $\tau$ of the variables studied here (for our
zero-stress conditions). In particular, $\tau$ for the in-plane area
$A_{\|}$ appreciably increases as the system size rises, and for the largest
sizes discussed here we have found autocorrelation times in the 
order of $10^5$ simulation steps.  

PIMD simulations can be used to study the atomic delocalization
at finite temperatures. This includes a thermal (classical) delocalization,
as well as a delocalization associated to the quantum character of the
atomic nuclei, which can be quantified by the extension of the paths
associated to a given atomic nucleus.
For each quantum path, one can define the center-of-gravity (centroid) as
\begin{equation}
   \overline{\bf r} = \frac{1}{L} \sum_{i=1}^L {\bf r}_i  \; ,
\label{centr}
\end{equation}
where ${\bf r}_i$ is the position of bead $i$ in the associated ring polymer.
Then, the mean-square displacement $(\Delta r)^2$ of the atomic nuclei
along a PIMD simulation run is defined as
\begin{equation}
  (\Delta r)^2 =  \frac{1}{L} \left< \sum_{i=1}^L
           ({\bf r}_i - \left< \overline{\bf r} \right>)^2
           \right>    \, ,
\label{deltar2}
\end{equation}
where $\langle ... \rangle$ indicates an ensemble average.

The kinetic energy of a particle is related to its quantum delocalization,
or in our context, to the spread of the paths associated to it,
which can be measured by the mean-square ``radius-of-gyration''
$Q_r^2$ of the ring polymers:
\begin{equation}
  Q_r^2 = \frac{1}{L} \left< \sum_{i=1}^L
             ({\bf r}_i - \overline{\bf r})^2 \right>    \, .
\label{qr2}
\end{equation}
A smaller $Q_r^2$ (higher particle localization) corresponds to a larger 
kinetic energy, in line with Heisenberg's uncertainty 
principle.\cite{gi88,gi90}

The total spatial delocalization of a quantum particle at a finite 
temperature includes, in addition to $Q_r^2$, another term which
takes into account motion of the centroid $\overline{\bf r}$, i.e. 
\begin{equation}
    (\Delta r)^2 = Q_r^2 + C_r^2  \, ,
\label{deltar2b}
\end{equation}
with
\begin{equation}
 C_r^2 =  \left< \left( \overline{\bf r} - \langle \overline{\bf r} \rangle
                \right)^2 \right>
       =  \langle  \overline{\bf r}^2 \rangle -
          \langle  \overline{\bf r} \rangle^2  \, .
\label{cr2}
\end{equation}
This term $C_r^2$ can be considered as a semiclassical thermal contribution
to $(\Delta r)^2$, since at high $T$ it converges to the mean-square
displacement given by a classical model, and each quantum path collapses onto 
a single point ($Q_r^2 \to 0$).

For our case of graphene, we call $(x, y)$ the coordinates on
the plane defined by the simulation cell, and $z$ the out-of-plane
perpendicular direction.  Then, we will have expressions similar to
those given above for each direction $x$, $y$, and $z$.
For example, $Q_r^2 = Q_x^2 + Q_y^2 + Q_z^2$ and 
$(\Delta z)^2 = Q_z^2 + C_z^2$.

\section{Results}

\subsection{Interatomic distances}

In this section we present results for interatomic distances in graphene.
The temperature dependence of the equilibrium C--C distance,
$d_{\rm C-C}$, as derived from our PIMD simulations at zero applied stress
is displayed in Fig.~1 (squares).
In the low-temperature limit, we find an interatomic distance of
1.4287(1) \AA, which increases for increasing temperature, as could
be expected from the thermal expansion of the graphene sheet.
We note that the size effect of the finite simulation cell is negligible
for our purposes. In fact, for a given temperature, we found no difference 
between the results for $d_{\rm C-C}$ obtained for the considered cell sizes,
i.e., differences between data for $N > 200$ were 
similar to the error bars obtained for each cell size (much less than
the symbol size in Fig.~1).

\begin{figure}
\vspace{-1.0cm}
\includegraphics[width= 8.5cm]{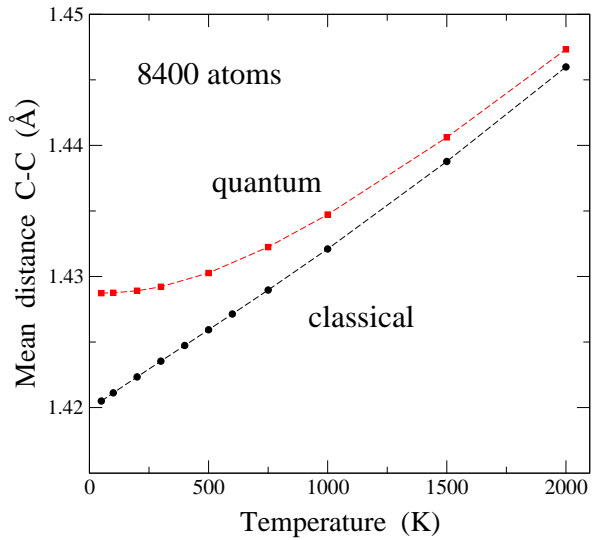}
\vspace{-0.5cm}
\caption{
Temperature dependence of the C--C distance for a simulation cell
containing 8400 atoms at zero external stress. Circles and squares
represent results of classical and PIMD simulations, respectively.
Dashed lines are guides to the eye.
Error bars are less than the symbol size.
}
\label{f1}
\end{figure}

For comparison with the results of the quantum simulations,
we also present in Fig.~1 the temperature dependence of $d_{\rm C-C}$ in 
the classical limit with the same LCBOPII potential, as derived from
MD simulations (circles). These results show at low $T$ a nearly 
linear increase,
similar to that found for lattice parameters of crystalline solids in 
a classical approximation, which is known to violate the third law of
thermodynamics\cite{ca60,wa72} (i.e., the thermal expansion coefficient should 
vanish for $T \to 0$). This anomaly of the classical model is remedied in 
the quantum simulations, which yield a vanishing slope for the $d_{\rm C-C}$
vs $T$ plot in the low-temperature limit.
The results of the classical simulations converge at low $T$ to an
interatomic distance of 1.4199 \AA, corresponding to the minimum
energy for a flat graphene sheet. 
This value is close to the distance $d_{\rm C-C}$ derived from 
{\em ab-initio} simulations in the limit $T \to 0$.\cite{po11b} 

The quantum simulations predict an interatomic distance larger than
the classical calculation, due basically to the fact that zero-point
motion of the carbon atoms in the quantum model detects the anharmonicity
of the interatomic potential even for $T \to 0$.
In this limit, we find a zero-point expansion of the C--C distance of
$8.8 \times 10^{-3}$ \AA, i.e., the bond distance increases by a
0.6\% respect the classical prediction.
This is close to a zero-point expansion of 0.53\% derived by 
Brito {\em et al.}\cite{br15} from path-integral Monte Carlo simulations.  
This increase in mean bond length is much larger than the precision
currently achieved in the determination of cell
parameters from diffraction techniques.\cite{ya94,ra93b,ka98}  
The bond expansion due to nuclear quantum effects decreases as 
temperature rises, as should happen, since in the high-$T$ limit the
classical and quantum predictions have to converge one to the other. 
We observe, however,
that at a temperature of 2000~K the C--C distance obtained from the quantum
simulations is still larger than the classical prediction
(the difference is much larger than the error bars).

In the quantum results,
the increase in interatomic distance from $T$ = 0~K to room temperature 
($T$ = 300~K) is small, and amounts to $\sim 5 \times 10^{-4}$ \AA, 
more than 15 times smaller than the zero-point expansion.
Note that the bond expansion associated to zero-point motion is
in the order of the thermal expansion predicted by the classical model
from $T$~=~0 to 800~K.
These results (classical and quantum) for a single layer of graphene
are qualitatively similar to those found earlier from simulations of
carbon-based materials. For diamond, in particular, it was found a zero-point
expansion of the lattice parameter $\Delta a = 1.7 \times 10^{-2}$ \AA,
a 0.5\% of the classical prediction.\cite{he00c}

The thermal bond expansion as well as the zero-point bond dilation
discussed here are a signature of the anharmonicity of the interatomic
potential, similar to 3D crystalline solids (see above). In the case of
graphene this effect is basically due to anharmonicity in the
stretching vibrations of the sp$^2$ C-C bond. A more involved 
anharmonicity appears in the description of the thermal variation
of the graphene area, due to the coupling between in-plane and
out-of-plane modes, as shown in the following section.

\subsection{Layer area}

The simulations (both classical MD and PIMD) presented here were carried
out in the isothermal-isobaric ensemble, as explained in Sec.~II.
This means that in a simulation run we fix the number of atoms $N$,
the temperature $T$, and the applied stress in the $(x, y)$ plane
($P = 0$ in our simulations), 
thus allowing changes in the area of the simulation cell 
on which periodic boundary conditions are applied.
Since carbon atoms are free to move in the out-of-plane direction
(coordinate $z$), it is obvious that in general any measure of
the ``real'' surface of the graphene sheet will give a value larger than
the area of the simulation cell in the $(x, y)$ plane.
Taking into account that the actual simulations yield directly the mean 
in-plane area $A_{\|}$ for given $N$, $T$, and $P$, and that this quantity 
has been discussed earlier in the literature, we present it for our 
classical and quantum simulations. 

\begin{figure}
\vspace{-1.0cm}
\includegraphics[width= 8.5cm]{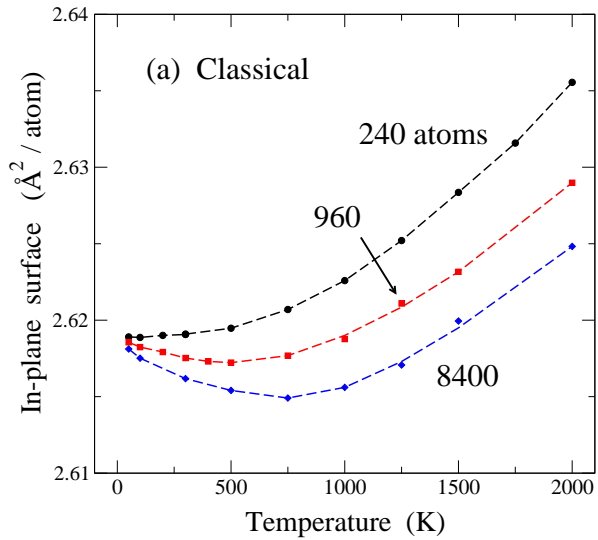}
~\vspace{-2.0cm}
\includegraphics[width= 8.5cm]{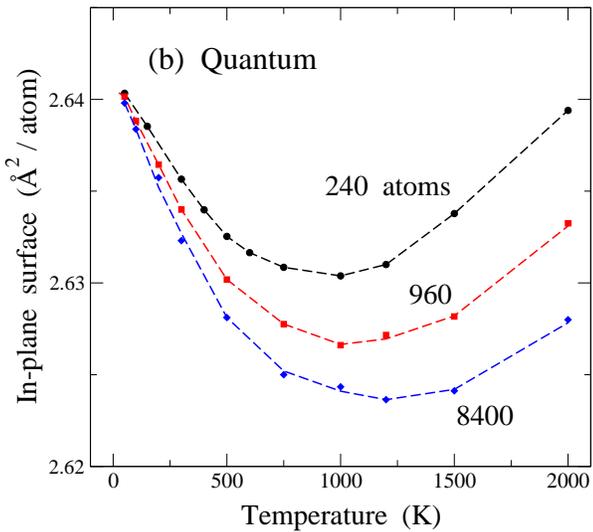}
\vspace{1.5cm}
\caption{
Temperature dependence of the mean in-plane surface $A_{\|}$.
(a) Results of classical MD simulations for three system sizes:
$N$ = 240 (circles), 960 (squares), and 8400 (diamonds).
(b) Results derived from PIMD simulations for the same system sizes
as in (a).  Dashed lines are guides to the eye.
}
\label{f2}
\end{figure}

In Fig.~2 we display the temperature dependence of the in-plane area 
$A_{\|}$ associated to the graphene 
layer, as derived from (a) classical MD and (b) PIMD simulations.
In each plot, results for three cell sizes are given: $N$ = 240 (circles),
960 (squares), and 8400 (diamonds) atoms.
Let us look first at the results of the classical simulations in 
Fig.~2(a). For the larger sizes (960 and 8400 atoms) one observes first
a decrease in $A_{\|}$ for rising $T$, and at higher temperatures $A_{\|}$ 
increases with $T$. For the smallest size presented here, the low-temperature
decrease in $A_{\|}$ is also present, although almost unobservable in the 
figure. Moreover, one sees that the normalized in-plane surface per atom
is smaller for larger simulation cells.
These results are similar to those found in earlier classical Monte Carlo 
and MD simulations of graphene single layers.\cite{za09,ga14,br15}
There are two competing effects which explain this temperature dependence
of $A_{\|}$, as discussed below.

In Fig.~2(b) we present the temperature dependence of $A_{\|}$, as derived
from PIMD simulations, for the same cell sizes as in Fig.~2(a).
This dependence is qualitatively similar to that corresponding to the
classical results, but in the quantum simulations the low-temperature
decrease of the in-plane area is more pronounced.
This is particularly visible for the smallest size presented here ($N$ = 240),
for which such a decrease in $A_{\|}$ for rising $T$ is almost inappreciable
in the classical results,
whereas it is clearly visible in the results of the quantum simulations.
It is important to note that, in spite of the large differences in the
in-plane area per atom for the different system sizes, in each case
(classical or quantum) all sizes converge at low $T$ to a single value.
This could be expected because for $T \to 0$ the graphene sheet becomes
strictly planar in the classical case and close to 
planar in the quantum 
model (a zero-point effect).

\begin{figure}
\vspace{-1.0cm}
\includegraphics[width=8.5cm]{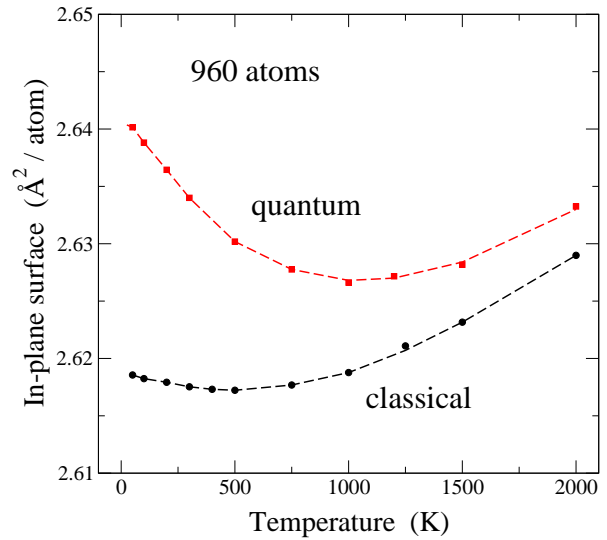}
\vspace{-0.5cm}
\caption{
Temperature dependence of the mean in-plane surface $A_{\|}$ for
$N$ = 960 atoms and zero external stress, as derived from classical
MD (circles) and PIMD simulations (squares).
Dashed lines are guides to the eye.
}
\label{f3}
\end{figure}

For a direct comparison of the in-plane area $A_{\|}$ obtained from
classical and quantum simulations, we show in Fig.~3 both sets of results
for system size $N$ = 960.
In the limit $T \to 0$, the difference between them converges to 
0.022 \AA$^2$. This difference decreases for rising temperature, 
as nuclear quantum effects become less relevant. 
For this system size, the overall behavior of $A_{\|}$ vs $T$ is similar
for both classical and quantum models, in the sense that $d A_{\|} / d T < 0$
at low temperature and this derivative becomes positive at high $T$.
However, the slope of the $A_{\|}$ vs $T$ curve at low temperature is much 
larger for the quantum than for the classical model. 

In view of the results shown in Fig.~3, it seems that $d A_{\|} / d T$
could not vanish for $T \to 0$, in disagreement with the third law of
thermodynamics\cite{ca60,wa72} (This apparent inconsistency is however
not real, as shown below in Sec.~IV).
Moreover, one can ask if $A_{\|}$ can be treated as a true extensive variable,
given that it shows a strong size effect, as shown in Fig.~2. In any case,
one can use $A_{\|}$ as a helpful variable to carry out simulations of
layered systems such as graphene, and indeed it is $A_{\|}$ which defines 
the area of the simulation cell in the $(x, y)$ plane. This in-plane area 
has been studied earlier in several works\cite{ga14,br15,za09,lo16,ch15} as 
a function of temperature and external stress, but one can argue that 
it is not necessarily a variable to which one can associate properties 
of a real material surface. 

This can be further illustrated by considering a ``real'' surface $A$
(in 3D space) for graphene, as can be defined from the actual bonds 
between carbon atoms. 
Thus, we define a mean area per atom from the mean C--C distance 
derived from the simulations, as $A = 3 \sqrt{3} \, d_{\rm C-C}^2 / 4$
(see Fig.~7 in Ref.~\onlinecite{ra16}).
A similar expression has been used by Hahn {\em et al.}\cite{ha16} 
to study nanocrystalline graphene from atomistic simulations. 
As the actual interatomic distance changes
from bond to bond in a simulation step and for each bond it fluctuates
along a simulation run, $A$ is a mean value for a given temperature,
and fluctuations of the instantaneous area depend on the fluctuations 
of the interatomic distance.
One could employ alternative definitions for the real area, based
for example on the areas of the hexagons that make up the graphene
sheet, or on a triangulation based on the atomic positions. 
We have checked that such definitions yield area values slightly 
smaller than the area $A$ defined above, but in the present conditions
(zero applied stress and not very large out-of-plane fluctuations)
this difference is not relevant for our current purposes. 

\begin{figure}
\vspace{-1.0cm}
\includegraphics[width=8.5cm]{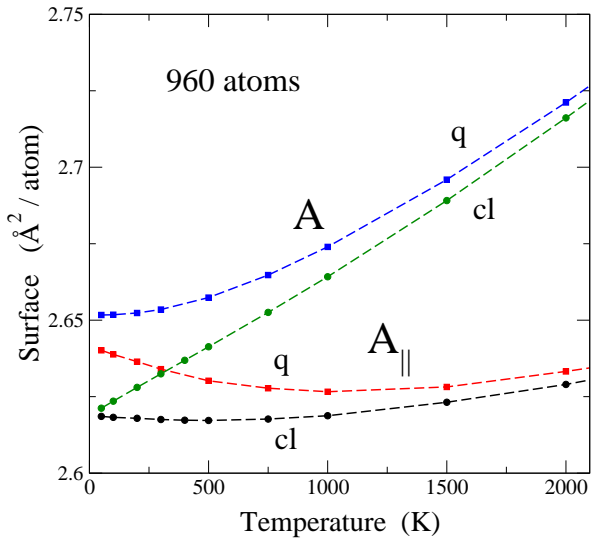}
\vspace{-0.5cm}
\caption{
Mean in-plane ($A_{\|}$) and 3D graphene surface ($A$) as a function
of temperature for $N$ = 960 atoms and zero applied stress.
In both cases circles and squares show results of classical (labeled as ``cl'')
and quantum simulations (labeled as ``q''), respectively.
Error bars are less than the symbol size.
Dashed lines are guides to the eye.
}
\label{f4}
\end{figure}

In Fig.~4 we show the areas $A$ and $A_{\|}$ vs temperature 
for a simulation cell including 960 atoms. In both cases, we present
results from classical and PIMD simulations.
The surface $A$ is larger than $A_{\|}$, and the difference between both
increases with temperature. In fact, $A_{\|}$ is the 2D projection
of $A$, and the actual surface becomes 
increasingly bent as temperature is
raised and out-of-plane atomic displacements are larger.
For the area $A$ we do not observe the anomalous decrease displayed 
by $A_{\|}$ in both classical and quantum cases at low temperature.
Moreover, the area $A$ derived from quantum simulations shows a temperature
derivative $d A / d T$ approaching zero as $T \to 0$, in agreement with 
low-temperature thermodynamics.

For the classical results, we find that both surfaces $A$ and $A_{\|}$
converge one to the other in the low-$T$ limit, as expected for a convergence
of the actual graphene surface to a plane when out-of-plane atomic 
displacements go to zero as $T \to 0$.
In the quantum case, $A$ and $A_{\|}$ become closer as temperature is reduced,
but in the low-$T$ limit $A$ is still larger than $A_{\|}$, and the 
extrapolated difference $A - A_{\|}$ amounts to 
$1.1 \times 10^{-2}$ \AA$^2$/atom for $T = 0$. 
This difference is due to the fact that even for $T \to 0$ the
graphene layer is not strictly planar due to zero-point 
motion of the carbon atoms in the out-of-plane direction.

We now discuss the behavior of $A_{\|}$ as a function of $T$, with
$d A_{\|} / d T < 0$ at relatively low temperature. 
There appears a competition between two opposite factors.
On one side, the actual area $A$ increases as temperature is raised,
as shown in Fig.~4 for both classical and quantum simulations.
On the other side, bending (rippling) of the surface 
causes a decrease in its 2D projection, i.e., in $A_{\|}$.
At low $T$, the decrease associated to out-of-plane motion dominates
the thermal expansion of the actual surface, and thus $d A_{\|} / d T < 0$.
This is particularly marked for the quantum results, because in this 
case the thermal expansion at low $T$ is very small.
At high temperature, the increase in $A$ dominates the reduction
in the projected area associated to out-of-plane atomic displacements.
For example, for the classical results, $A$ increases roughly linearly
with $T$, but the mean-square atom displacements (causing the reduction of
$A_{\|}$) grow up slower than linearly, as a consequence of anharmonicity.
This is in line with the discussion given by Gao and Huang\cite{ga14}
for the thermal behavior of $A_{\|}$ observed in classical MD simulations 
of a single graphene layer, as well as with the trends for $d A_{\|} / d T$
calculated by Michel {\em et al.}\cite{mi15b}

The discussion of a real area $A$ and a projected area $A_{\|}$ presented
here is reminiscent of the analysis carried out earlier in the context of
biological membranes, where consideration of a ``true'' area is 
important to describe some properties of those membranes.\cite{ch15}
In fact, it has been shown that values of the membrane compressibility may
be very different when they are related to $A$ or to $A_{\|}$.
Something similar can happen for graphene, and this topic requires
further research. 

To summarize the results of this section, we emphasize that changes in 
interatomic distances and in the graphene area (both $A$ and $A_{\|}$)
are important anharmonic effects. On one side, changes in the distance
$d_{\rm C-C}$ and in the real area $A$ are mainly due to anharmonicity
of the C-C stretching vibration. The dilation of the C-C bond 
displayed in Fig.~1 equally appears in a strictly planar graphene, and 
shows a temperature dependence similar to that of interatomic distances in
solids. In real graphene, $d_{\rm C-C}$ will be also affected by 
out-of-plane vibrations, which in fact couple with in-plane vibrational 
modes, but this coupling changes $d_{\rm C-C}$ less than the only 
anharmonicity of the in-plane modes, at least for the conditions considered 
here. Something different occurs for the projected area $A_{\|}$, as
manifested by the strong size effect which it displays.
In this case, the coupling between in-plane and out-of-plane modes is
crucial to understand the temperature dependence of $A_{\|}$ in the
absence of an external stress ($P = 0$). The amplitude of 
out-of-plane vibrations strongly depend on the size $N$,\cite{ga14} 
causing the large size effect in $A_{\|}$. 
At low temperatures, all these anharmonic effects are clearly
enhanced by quantum motion, as shown for example in Fig.~3 for the
projected area. This is due to the fact that anharmonicity is manifested
in the quantum model even in the limit $T \to 0$, contrary to the classical
case, where it gradually appears as temperature is raised.
This is further quantified and discussed in the next section.

\subsection{Vibrational energy}

In the zero-temperature limit we find in the classical approach
a strictly planar graphene surface with an 
interatomic distance
of 1.4199 \AA.  This corresponds to fixed atomic nuclei on the
equilibrium sites without spatial delocalization, defining a minimum 
energy $E_0$, taken as a reference for our calculations.
In the quantum approach, the low-temperature limit includes out-of-plane
atomic fluctuations due to zero-point motion, so that the graphene layer
is not strictly planar, as commented above. 
Moreover, anharmonicity of in-plane vibrations
cause a slight zero-point bond expansion, giving an interatomic distance
of 1.4287 \AA, as shown in Fig.~1. 

We now turn to the results of our simulations at finite temperatures,
and will discuss the resulting energy for graphene.
The internal energy, $E(T)$, at temperature $T$ can be written as:
\begin{equation}
    E(T) =  E_0 + E_{\rm el}(A) + E_{\rm vib}(A,T)   \, ,
\label{et}
\end{equation}
where $E_{\rm el}(A)$ is the elastic energy corresponding to an
area $A$, and $E_{\rm vib}(A,T)$ is the vibrational energy of the system.
Our simulations directly yield $E(T)$ in both classical and quantum approaches.
We can then split the internal energy $E(T) - E_0$ into
an elastic and a vibrational part.

To define the elastic energy corresponding to an area $A$, we
have carried out calculations of the (classical) energy of a
supercell including 960 atoms, expanding it isotropically and
keeping it flat. This strict 2D atomic layer gives us a reference
to evaluate the  vibrational energy.  The elastic energy
$E_{\rm el}(A)$ increases with $A$, and for small expansion it can be
approximated as
$E_{\rm el}(A) \approx K (A - A_0)^2$, with $K$ = 2.41 eV/\AA$^2$.
For the area $A$ derived from our classical and PIMD simulations at
300~K, we found elastic energies of 0.4 and 2.8 meV/atom, 
respectively. This much larger value for the quantum approach
is a consequence of the larger surface $A$, as compared to the classical 
limit at $T$ = 300~K.
Those values of the elastic energy increase to 20.8 and 
22.9~meV/atom at 2000~K.

\begin{figure}
\vspace{-1.0cm}
\includegraphics[width=8.5cm]{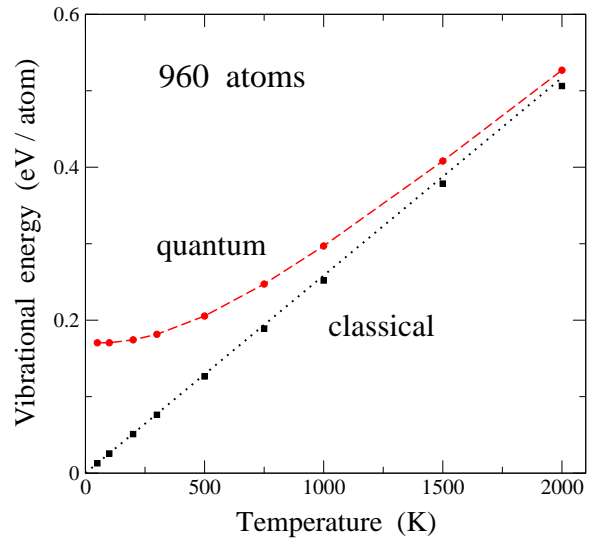}
\vspace{-0.5cm}
\caption{
Temperature dependence of the vibrational energy per carbon atom, as
derived from classical (squares) and PIMD simulations (circles) of graphene.
The simulation cell included 960 atoms.
The dotted line corresponds to the classical limit of the vibrational
energy per atom in a harmonic approximation: $E_{\rm vib}^{\rm cl} = 3 k_B T$.
The dashed line is a guide to the eye.
Error bars are less than the symbol size.
}
\label{f5}
\end{figure}

Thus, we obtain the vibrational energy $E_{\rm vib}(A,T)$ from the
results of our simulations by subtracting in each case the elastic 
energy from the internal energy $E(T)$.
 In Fig.~5 we show the temperature dependence of $E_{\rm vib}(A,T)$,
derived from simulations for a supercell including 960 carbon atoms.
At 300~K, the vibrational energy amounts to 182 meV/atom, 
somewhat smaller than the values reported for diamond of 195 and
210 meV/atom, yielded by path-integral simulations with Tersoff-type
and tight-binding potentials, respectively.\cite{he00c,ra06} 
This has to be compared to the value expected in a classical harmonic
approximation: $E_{\rm vib}^{\rm cl}(T) = 3 k_B T$ = 77.6 meV/atom
($k_B$ is Boltzmann's constant).
Note also that the elastic energy $E_{\rm el}$ in the quantum approach
at 300~K is 1.5\% of the internal energy $E(T) - E_0$, most of this
energy corresponding to the vibrational energy $E_{\rm vib}$.

\begin{figure}
\vspace{-1.0cm}
\includegraphics[width=8.5cm]{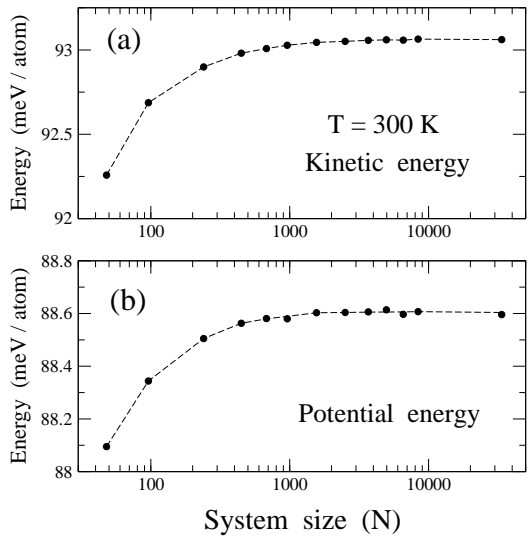}
\vspace{-0.5cm}
\caption{
Size dependence of the vibrational kinetic [panel (a)] and potential
energy of graphene [panel (b)], as derived from PIMD simulations at 300~K.
Dashed lines are guides to the eye.
For the kinetic and potential energy, error bars are smaller and
in the order of the symbol size, respectively.
}
\label{f6}
\end{figure}

PIMD simulations  allow us to separate the potential ($E_p$) and kinetic
($E_k$) contributions to the vibrational energy,\cite{he82,gi88b,gi90}
so that $E_{\rm vib} = E_k + E_p$.
In the classical limit, the kinetic energy amounts to $3 k_B T / 2$ per atom. 
In the quantum case, however, $E_k$ depends on the
spatial delocalization $Q_r^2$ of the quantum paths, which can be 
obtained from the simulations as indicated above [see Eq.~(\ref{qr2})].
In our simulations of graphene, both kinetic and potential energy were
found to slightly increase with system size, but their convergence is
rather fast.
In Fig.~6, we present $E_k$ and $E_p$ derived from PIMD simulations
as a function of cell size at $T$ = 300~K.
Symbols indicate results of our simulations for $E_k$ [panel (a)]
and $E_p$ [panel (b)], whereas dashed lines are guides to the eye.
In both $E_k$ and $E_p$ there 
appears a shift of less than 1 meV/atom when increasing the cell
size from 40 atoms to the largest size considered here (about 30,000
atoms). For cells in the order of 1000 atoms the size effect is almost
inappreciable when compared to the largest cell.  
The potential energy is found to be smaller 
than the kinetic energy, indicating an appreciable anharmonicity of the
lattice vibrations (see below).

\begin{figure}
\vspace{-1.0cm}
\includegraphics[width=8.5cm]{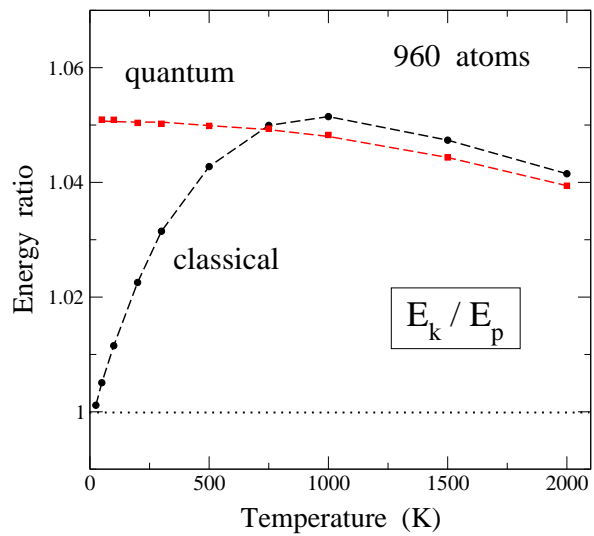}
\vspace{-0.5cm}
\caption{
Temperature dependence of the ratio $E_k / E_p$ between
vibrational kinetic and potential energy obtained from
classical (circles) and PIMD (squares) simulations.
These results were obtained for a simulation cell including
960 atoms.
Dashed lines are guides to the eye.
The dotted line shows the ratio $E_k / E_p = 1$ expected for
pure harmonic vibrations.
}
\label{f7}
\end{figure}

Differences between the kinetic and potential contribution to the
vibrational energy can be used for a quantification of the
anharmonicity in condensed matter, as has been discussed earlier from
path-integral simulations, e.g., for H impurities in silicon\cite{he95} 
and van der Waals solids.\cite{he02} 
In Fig.~7 we display the ratio $E_k / E_p$ between kinetic and potential
parts of the vibrational energy of graphene as a function of temperature.
Results are shown for classical (circles) and PIMD (squares) simulations
carried out for a system size $N$ = 960 atoms.
The effect of system size on the energy ratio is a slight shift in both
cases, classical and quantum, as can be derived from results such as 
those presented in Fig.~6. 

For a purely harmonic model for the vibrational modes, one expects
$E_k = E_p$ (virial theorem\cite{la80,fe72}), i.e., an energy ratio equal 
to unity at any temperature in both classical and quantum approaches.
The ratio $E_k / E_p = 1$ is found in the classical model for
the low-temperature limit, since in this case the atomic motion does
not feel the anharmonicity of the interatomic interactions, due to
the vanishingly small vibrational amplitudes.
This is not the case of the quantum results for $T \to 0$, because
now the vibrational amplitudes remain finite and feel the anharmonicity.
In particular, we find $E_k > E_p$, and at low temperature this ratio
amounts to about 1.05.
In the classical model, $E_k / E_p$ increases at low $T$, due to
an enhancement of vibrational amplitudes, as 
in this approach the mean-square atomic displacements increase
at low $T$ linearly with rising temperature.
This increase in the energy ratio continues until about $T$ = 1000~K,
and at higher temperatures $E_k / E_p$ is found to decrease, close
to the quantum result.
This slow decrease for rising $T$ obtained for the energy ratio from 
both classical and PIMD simulations does not mean that the difference
$E_k - E_p$ decreases (in fact it rises with $T$), but is due to the 
large increase in the denominator $E_p$.
Also, for $T \gtrsim$ 1000~K the ratio $E_k / E_p$ turns out to
be larger for the classical case than in the quantum approach, even
though the difference $E_k - E_p$ is similar in both cases,
because in the classical approach $E_p$ is smaller.

Turning to the energy results for the quantum approach at $T \to 0$,
it is interesting to note that earlier analysis of anharmonicity
in solids, based on quasiharmonic approximations and perturbation
theory indicate that low-temperature changes in the vibrational energy
with respect to a harmonic calculation are mainly due to the kinetic energy.
In fact, assuming  perturbed harmonic oscillators (with perturbations
of $x^3$ or $x^4$ type) at $T = 0$, the first-order change in the
energy is caused by changes in $E_k$, the potential
energy remaining unaltered with respect to its unperturbed
value.\cite{la65,he95}

\subsection{Atomic motion: Out-of-plane displacements}

In this section we present results for the mean-square displacements
of carbon atoms in graphene, as obtained from our PIMD simulations.
We mainly focus on the character of these atomic displacements, 
i.e., if they can be described by a classical model, or the C atoms 
mainly behave as quantum particles. One obviously expects that 
the quantum description will become more relevant as the temperature
is lowered, but we show that the system size also plays an important role
in this question.  We use the notation introduced in Sec.~II.

\begin{figure}
\vspace{-1.0cm}
\includegraphics[width=8.5cm]{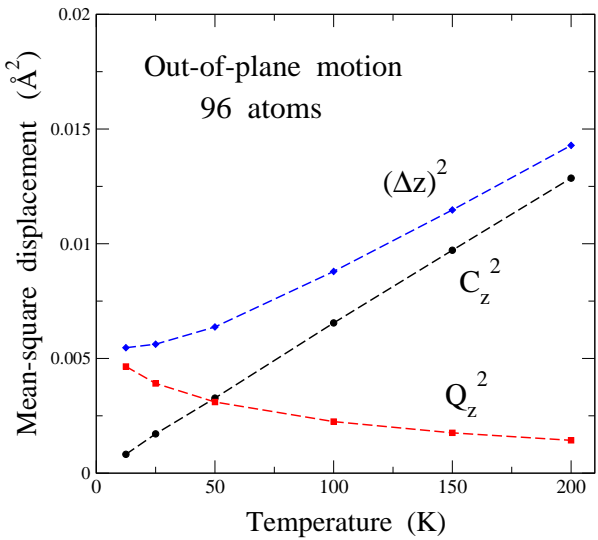}
\vspace{-0.5cm}
\caption{
Temperature dependence of the mean-square displacement along
the out-of-plane direction, $(\Delta z)^2$  (diamonds),
which is the sum of $Q_z^2$ (squares) and $C_z^2$ (circles).
These results were obtained from PIMD simulations for a
graphene cell containing 96 atoms.
Dashed lines are guides to the eye.
}
\label{f8}
\end{figure}

In Fig.~8 we display results for the motion in the out-of-plane
direction $z$, derived for a cell including 96 atoms. 
Diamonds represent the mean-square displacement
$(\Delta z)^2$ as a function of temperature [see Eq.~(\ref{deltar2})]. 
As explained above, this displacement can
be split into two parts: one of them properly quantum in nature, 
corresponding to the spread of the quantum paths, $Q_z^2$ (squares),
and another of semiclassical character, taking into account the centroid 
motion, i.e., global displacements of the paths, $C_z^2$ (circles).
In the limit $T \to 0$, $C_z^2$ vanishes and $Q_z^2$ converges to a value 
of about $6 \times 10^{-3}$ \AA$^2$. 
$Q_z^2$ decreases for increasing temperature, whereas $C_z^2$ increases
almost linearly with $T$. For the system size shown in Fig.~8 (N = 96), 
both contributions to $(\Delta z)^2$ are nearly equal at $T$ = 50~K. 
At higher temperatures, the semiclassical part $C_z^2$ is the main 
contribution to the atomic displacements on the out-of-plane direction.
Values of $C_z^2$ obtained from our PIMD simulations coincide within error
bars with the mean-square atomic displacement in the $z$ direction obtained 
from classical MD simulations.
The proper quantum delocalization can be estimated from the mean extension 
of the quantum paths in the $z$ direction, i.e., from $Q_z^2$. 
At 25 and 300~K we find an average 
extension $(\Delta z)_Q = (Q_z^2)^{1/2} \approx$ 0.06 and 0.03 \AA, 
respectively.

The vibrational amplitudes in the layer plane $(x, y)$ are smaller than
in the $z$ direction. This is true for both, quantum and classical 
contributions. In the limit $T \to 0$, zero-point motion yields a
mean-square displacement on the plane 
$Q_{\|}^2 = 3.6 \times 10^{-3}$ \AA$^2$, which means 
$Q_x^2$ = $Q_y^2$ = $1.8 \times 10^{-3}$ \AA$^2$. Thus, for each direction 
in the $(x, y)$ plane we find a value about three times smaller than
the zero-point mean-square displacement in the $z$ direction.
This is due, of course, to the low-frequency out-of-plane modes, which
cause a larger vibrational amplitude.
At finite temperatures, the difference between vibrational amplitudes
in-plane and out-of-plane are larger than at $T = 0$.
At 300~K, we find $(\Delta z)^2 / (\Delta x)^2$ = 11.9 and 61.8 
for simulation cells with 96 and 960 atoms, respectively.
This is a consequence of the appearance of long-wavelength modes
(with low frequency) with larger vibrational amplitudes in the
larger cells, as explained below.

\begin{figure}
\vspace{-1.0cm}
\includegraphics[width=8.5cm]{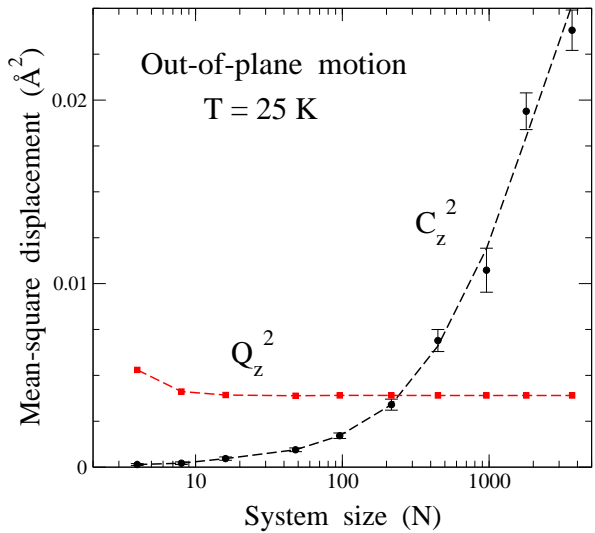}
\vspace{-0.5cm}
\caption{
Size dependence of the mean-square displacements
$C_z^2$ (classical, circles) and $Q_z^2$ (quantum, squares)
along the out-of-plane direction.
These data were obtained from PIMD simulations at $T$ = 25~K.
Dashed lines are guides to the eye.
Error bars for $Q_z^2$ are smaller than the symbol size.
}
\label{f9}
\end{figure}

The picture shown in Fig.~8 for the atom displacements in the 
$z$ direction is qualitatively the same for different system 
sizes, but the temperature region where $Q_z^2$ or $C_z^2$ is the main 
contribution to $(\Delta z)^2$ is strongly dependent on 
the system size 
$N$. This is mainly due to the enhancement of the classical-like contribution
for increasing size, a fact already observed and described earlier
from results of classical MD simulations.\cite{lo09,ga14,ra16}
In Fig.~9 we present mean-square displacements $Q_z^2$ and $C_z^2$ in
the out-of-plane direction at $T$ = 25~K for different $N$.
One observes that the quantum contribution $Q_z^2$ converges fast as $N$
is increased, and in fact for $N > 50$, changes in $Q_z^2$ are much smaller
than the symbol size. Clear finite-size effects are only observed for very 
small simulation cells.
The semiclassical contribution $C_z^2$ is small for small $N$, and grows 
with system size, so that at $T$ = 25~K it becomes similar to $Q_z^2$ 
for $N \sim 200$.
Calling $N_c$ the system size at which $C_z^2$ and $Q_z^2$ are equal, we
found that $N_c$ decreases for rising temperature, and for 50~K we have
$N_c \sim 100$ [see Fig.~8].

\begin{figure}
\vspace{-1.0cm}
\includegraphics[width=8.5cm]{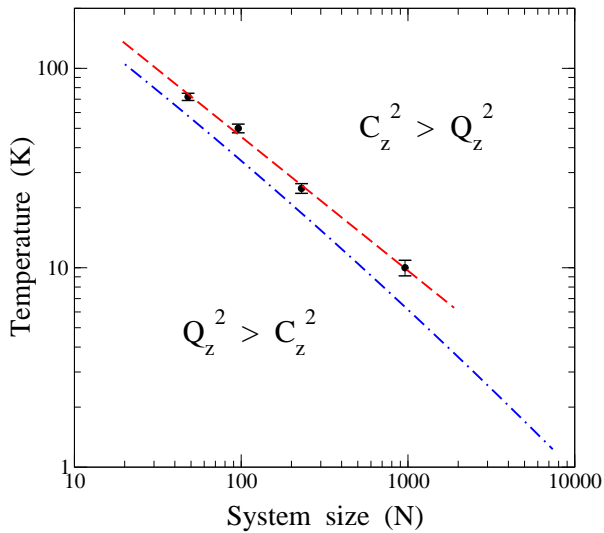}
\vspace{-0.5cm}
\caption{
$N - T$ plane showing the crossover from the region dominated by
quantum delocalization ($Q_z^2 > C_z^2$, below the dashed line) to
the region dominated by classical-like motion ($C_z^2 > Q_z^2$,
above the dashed line).
Data points were obtained from PIMD simulations for several system sizes.
The dashed line was obtained as a linear fit of $\log T$ vs $\log N$ for the
crossover temperatures derived from the simulations.
The dashed-dotted lined was derived from a harmonic model
[Eqs.~(\ref{cz2}) and (\ref{qz2b})] as explained in the text.
}
\label{f10}
\end{figure}

On the other side, for a given system size $N$, the ratio $Q_z^2 / C_z^2$ 
decreases for increasing $T$, and there is a crossover temperature $T_c$ 
for which this ratio is unity. For $T > T_c$ classical-like motion dominates 
the atomic mean-square displacement in the $z$ direction.
In Fig.~10 we show $T_c$ as a function of the system size $N$. 
Symbols are results derived from our PIMD simulations.
The dashed line through the data points was obtained as a linear fit of 
$\log T$ vs $\log N$ for the crossover temperatures derived from the 
simulations for several cell sizes.  
This is consistent with a power-law dependence of the crossover
temperature on system size, i..e, $T_c = a \, N^{-b}$ with a factor 
$a = 1002$ K and an exponent $b = 0.67$.

This means that for a given temperature (even very low), there is a
system size for which classical-like motion along the out-of-plane direction
dominates over atomic quantum delocalization.  Thus, extrapolating
the dashed line in Fig.~10, we find for $T$ = 1~K and 0.1~K 
crossover sizes $N_c \approx 3 \times 10^4$ and $9 \times 10^5$, respectively.
These values are much larger than the simulation cells which can be dealt
with in PIMD simulations at low temperatures. 
We note, however, that the dependence of
$T_c$ on $N$ may depart from the power-law given above (dashed line in 
Fig.~10) for larger system sizes. This is related to the size dependence of
$C_z^2$, which has been proposed to increase logarithmically with $N$
(from an atomistic model for the spectral amplitudes\cite{ga14,ra16})
or as a power-law (based on a scaling theory in the continuum medium
approach\cite{lo09}). In any case, this difference would be detected for 
system sizes much larger than those considered here in the PIMD simulations.

This can be rationalized in terms of vibrational modes in the $z$
direction appearing for different system sizes.
The mean-square displacements $C_z^2$ and $Q_z^2$ can be estimated in 
a harmonic approximation in the following way. 
The classical part is given by:
\begin{equation}
 C_z^2 \approx \frac1N \sum_{\bf k} \frac{k_B T}{m \omega({\bf k})^2} \, ,
\label{cz2}
\end{equation}
where $m$ is the atomic mass and the sum is extended to the wavevectors 
${\bf k} = (k_x, k_y)$ in the reciprocal lattice corresponding to the
two-dimensional simulation cell.\cite{ra16} 
This expression is valid for relatively low temperatures,
as for high $T$ anharmonic effects cause $C_z^2$ to
rise sublinearly with $T$.\cite{ga14}
On the other side, in the limit $T \to 0$, $Q_z^2$ converges to
\begin{equation}
 Q_z^2(0) \approx \frac1N \sum_{\bf k} \frac{\hbar}{2 m \omega({\bf k})} \, .
\label{qz2}
\end{equation}
Increasing the system size $N$ causes the appearance of vibrational modes
with longer wavelength $\lambda$. In fact, one has an effective cut-off 
$\lambda_{max} \approx l$, with $l = (N A_{\|})^{1/2}$. 
Calling $k = |{\bf k}|$, this translates into 
$k_{min} = 2 \pi / \lambda_{max}$, i.e., $k_{min} \sim N^{-1/2}$.
We consider a dispersion relation $\omega({\bf k})$ of the form
$\rho \, \omega^2 = \sigma k^2 + \kappa k^4$,
consistent with an atomic description of graphene\cite{ra16}
($\rho$, surface mass density; $\sigma$, effective stress; 
$\kappa$, bending modulus).
Then, the sum in Eq.~(\ref{cz2}) diverges in the large-size limit due to
the $\omega({\bf k})^2$ term in the denominator and the two-dimensional 
character of the wavevector ${\bf k}$\cite{ga14,ra16}
(see results for $C_z^2$ vs $N$ at 25~K in Fig.~9).
However, the sum in Eq.~(\ref{qz2}) converges to a finite value for
large $N$, in agreement with the results of our PIMD simulations at
low temperature (see results for $Q_z^2$ vs $N$ in Fig.~9).

Given a temperature $T$, modes with frequency 
$\omega \lesssim  \omega_c(T) = k_B T / \hbar$ may be considered 
in the classical regime.
Then, increasing $N$ one reaches a system size $N_c(T)$ for which any further 
increase introduces new modes with frequency $\omega < \omega_c$, and 
therefore contributing
to increase the classical displacement $C_z^2$ more than the
quantum one $Q_z^2$.
Thus, at any finite temperature classical-like displacements will
dominate over quantum delocalization in the out-of-plane direction, provided
that the system size is larger than the corresponding $N_c$.

All this can be visualized in terms of the harmonic model.
Taking into account that at low temperatures $\sigma \approx 0$, the
dispersion relation in the $z$ direction is
$\omega({\bf k})^2 \approx \kappa k^4 / \rho$ (see Ref.~\onlinecite{ra16}).
At temperature $T$, the mean-square displacement can be approximated as
\begin{equation}
 (\Delta z)^2 \approx \frac1N \sum_{\bf k} \frac{\hbar}{2 m \omega({\bf k})} 
       \coth \left( \frac{\hbar \omega({\bf k})}{2 k_B T}  \right)  \, .
\label{qz2b}
\end{equation}
Then, from Eqs.~(\ref{cz2}) and (\ref{qz2b}), and remembering that 
$Q_z^2 = (\Delta z)^2 - C_z^2$, we calculate for each $N$ the temperature
at which $Q_z^2 = C_z^2$. This yields the curve shown as a dashed-dotted
line in Fig.~10, displaying a dependence of $T_c$ on $N$ very similar to
that derived from PIMD simulations for system sizes accessible in our 
calculations. In fact, for $N \lesssim 1000$, the results can be approximated
in both cases by a power-law with nearly the same exponent. For larger 
sizes, the harmonic calculation deviates to temperatures somewhat smaller
than the extrapolation of the simulation data.  
At this point, we cannot assure if this deviation is a real trend of the
physical system or only a consequence of the harmonic approach.
The main limitations of this approximation are the neglect
of anharmonicity in the vibrational modes (expected to be reasonably 
small at low $T$) and the overestimation of
frequencies in the ${\bf k}$-space region far from the $\Gamma$ point
({\bf k} = 0). In that region, $\omega$ increases with $k$ slower than
$k^2$ (see Ref.~\onlinecite{mo05,mi15,ra16}). 
Even with these limitations the harmonic 
model captures qualitatively, and almost quantitatively, the basic aspects 
of the competition between classical-like and proper quantum dynamics
of the carbon atoms in the $z$ direction.

We finally note that the competition between $C_z^2$ and $Q_z^2$ as a 
function of $N$ discussed here appears only in the $z$ direction. 
For motion in the $(x, y)$ plane, both $C_x^2$ and $Q_x^2$ converge fast 
with increasing system size to their asymptotic value, similarly
to the behavior known for vibrational motion in 3D solids.\cite{he14}

\section{Thermodynamic consistency}

Here we analyze the thermodynamic consistency of the 
results of our PIMD simulations. In particular, we discuss their
compatibility with the third law of thermodynamics.
According to this law, some variables such as the heat capacity
or thermal expansion should vanish in the limit $T \to 0$.\cite{ca60,wa72}

Concerning the heat capacity $c_P = d E / d T$ derived from the PIMD
simulations, we find for all system sizes considered here low-temperature
values compatible with the third law, i.e. $c_P \to 0$ for $T \to 0$.
This can be visualized in Fig.~5, where the vibrational energy obtained
from PIMD simulations fulfills $d E_{\rm vib} / d T \to 0$ at low $T$.
Something similar happens for the elastic energy $E_{\rm el}$, so that
the temperature derivative of the internal energy $E$ [see Eq.~(\ref{et})]
also vanishes at low $T$. This is similar to the results of earlier
path-integral simulations of 3D solids,\cite{he00c,he14} where the resulting
data were in agreement with the basic laws of thermodynamics. 

A more subtle question appears for the thermal expansion of graphene.
We first note that, following our definitions of 3D area $A$ and 
in-plane area $A_{\|}$, we consider two different definitions for the
areal thermal expansion coefficient: 
\begin{equation}
   \alpha = \frac{1}{A}  \left( \frac{\partial A}{\partial T} \right)_P
\end{equation}
and
\begin{equation}
 \alpha_{\|} = \frac{1}{A_{\|}}  
         \left( \frac{\partial A_{\|}}{\partial T} \right)_P   \, .
\end{equation}
The 3D area $A$ derived from our PIMD simulations displays a negligible 
size effect, as indicated in Sec.~III.B. Therefore, the same happens 
for the coefficient $\alpha$,
which vanishes in the zero-temperature limit and is found to be positive 
at all finite temperatures considered here (see Fig.~4). 

Concerning the in-plane area $A_{\|}$, we have shown that there appears
a strong size effect in both classical and quantum results (see Fig.~2).
Moreover, for both kinds of simulations the data shown in Figs.~2 and 3
seem to indicate that the in-plane thermal expansion coefficient 
$\alpha_{\|}$ is negative at low temperature, irrespective of the cell size.
This causes no particular problems for the classical results, as discussed
above. However, in the quantum case, one should expect a vanishing 
$\alpha_{\|}$ at temperatures low enough to have all vibrational modes 
(nearly) in their ground state, so that the system is ``frozen'' and 
no change in $A_{\|}$ could occur. 

\begin{figure}
\vspace{-1.0cm}
\includegraphics[width=8.5cm]{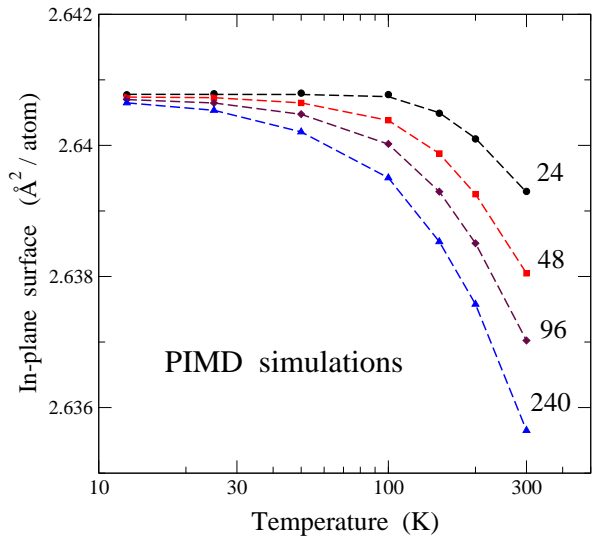}
\vspace{-0.5cm}
\caption{
Temperature dependence of the in-plane area $A_{\|}$ for several cell
sizes, as derived from PIMD simulations.
From top to bottom, $N$ = 24, 48, 96, and 240.
Dashed lines are guides to the eye.
}
\label{f11}
\end{figure}

A hint to solve this apparent inconsistency can be obtained from the 
discussion in Sec.~III.D concerning the vibrational modes appearing 
effectively for each cell size. The relevant modes for this purpose are 
the low-frequency out-of-plane modes (in the $z$ direction), since in-plane 
modes have much larger frequencies.
The results summarized in Fig.~10 indicate that the temperature region
where proper quantum motion dominates over classical-like motion in the
$z$ direction is strongly dependent on the system size. 
Thus, for
$N$ = 240 (the smallest size shown in Fig.~2), one has to go to
$T < 20$ K to observe the mode ``freezing'' mentioned above. 

Since the temperature for this ``freezing'' to occur increases
by reducing the cell size, making it more readily appreciable, we
have carried out some PIMD simulations for cell sizes and temperatures 
down to $N$ = 24 and $T$ = 12 K.
The results for $A_{\|}$ are displayed in Fig.~11 in a semilogarithmic
plot, including system sizes of 24, 48, 96, and 240 atoms.
In all cases we find that the $A_{\|}(T)$ curve becomes flat at low
temperature, as expected for a vanishing of $\alpha_{\|}$, but this
is more manifest for smaller sizes, in agreement with Fig.~10.
Moreover, $A_{\|}$ converges to the same value for the different
cell sizes: $A_{\|}(0)$ = 2.6408 \AA$^2$/atom.
For $N$ = 240 one observes that $A_{\|}(T)$ saturates at low $T$
to the same value as the smaller cell sizes, although this is
almost inappreciable at the scale of Fig.~2(b).

These low-temperature data for $A_{\|}(T)$ are consistent with the
trend $\alpha_{\|} \to 0$ in the low-temperature limit derived 
by Amorim {\em et al.},\cite{am14} who investigated the thermodynamic 
properties of 2D crystalline membranes using a first-order perturbation 
theory and a one-loop self-consistent approximation.
These authors also found that the limits $N \to \infty$ and $T \to 0$ 
do commute, a fact compatible with the results shown in Fig.~11, 
i.e., at low $T$ all system sizes yield the same results.
We find, however, that an evaluation of the low-temperature (quantum)
properties becomes harder in both limits. Going to larger $N$ does not
only mean to increase the system size, but to reduce the temperature 
to reach the proper quantum regime, with the consequent increase in
the Trotter number appearing in PIMD simulations.

\section{Summary}

In this paper we have presented results of PIMD simulations of a graphene
monolayer in the isothermal-isobaric ensemble at several temperatures and
zero external stress.
The relevance of quantum effects has been assessed by comparing results
given by PIMD simulations with those yielded by classical MD simulations.
Structural variables are found to change when quantum nuclear motion is
taken into account, especially at low temperatures.
Thus, the sheet area and interatomic distances change appreciably
in the range of temperatures considered here.

The LCBOPII potential model is known to describe fairly well various
structural and thermodynamic properties of carbon-like materials, and 
graphene in particular.  Here we have investigated
its reliability to study effects related to the quantum character of 
atomic nuclei, and their vibrational motion in particular.
Given the ability of this effective potential to describe rather
accurately the vibrational frequencies of graphene, one expects that
quantum effects associated to vibrational motion should be equally
described in a reliable manner by PIMD simulations using this potential.
The results obtained in the simulations have allowed us to propose
a consistent interpretation of the in-plane ($A_{\|}$) and ``real''
($A$) graphene surfaces.
In order to check these finite-temperature results,
it would be desirable to study structural and thermodynamic properties
of graphene from an {\em ab-initio} method.
This is, however, not feasible at present for the large supercells
required to study these properties.

Particular emphases has been laid on the atomic vibrations along the
out-of-plane direction.
Even though quantum effects are present in these vibrational modes, 
we have shown that at any finite temperature classical-like motion 
dominates over quantum delocalization, provided that the system size is 
large enough. This size effect is present in the quantum simulations
at low temperatures, as a consequence of the appearance of vibrational modes
with smaller wavenumbers in larger graphene cells.
Moreover, by comparing the kinetic and potential energy derived from PIMD
simulations, we have shown that vibrational modes display a nonnegligible 
anharmonicity, even for $T \to 0$.
This comparison indicates that the overall kinetic energy is larger
than the vibrational potential energy by about 5\% at low temperatures.

An important question related to the overall consistency of the simulation
results is their agreement with the basic principles of thermodynamics. 
This is usually taken for granted in the standard simulation methods
used today, but in the case of graphene some subtle questions may appear
due to its 2D character in a 3D world.
In particular, the third law of thermodynamics has to be satisfied, 
in the sense that proper thermodynamic variables should display at
low temperature a behavior compatible with this law, e.g., thermal expansion
coefficients should converge to zero for $T \to 0$. 
We have shown that this requirement is fulfilled by both the in-plane 
area $A_{\|}$ and the 3D area $A$ derived from PIMD simulations.

In summary, we have shown that 
(1) the so-called thermal contraction of graphene mentioned in the 
literature is in fact a reduction of the projected area $A_{\|}$ due 
to out-of-plane vibrations, and not a to an actual decrease in the real
area $A$ of the graphene sheet.
(2) The difference between $A$ and $A_{\|}$ increases for rising $T$
due to the larger amplitude of those vibrations.
(3) Zero-point expansion of the graphene layer due to nuclear quantum 
effects is not negligible, and it amounts to an increase of about 1\% in
the area $A$.
(4) The temperature dependence of the projected area $A_{\|}$ may be
qualitatively different when derived from classical or PIMD simulations,
even at temperatures between 300 and 1000 K (see Fig.~3).
(5) Anharmonicity of the vibrational modes is appreciable and should be 
taken into account in any finite-temperature calculation of the properties 
of graphene.
(6) The temperature region where out-of-plane vibrational modes are
predominantly of quantum character decreases as the system size rises.
This is important for a comparison of results derived from classical and 
quantum models for different system sizes. 
(7) Thermodynamic consistency of the results of PIMD simulations at low 
$T$ has been shown.

Path-integral simulations similar to those presented here may help
to understand low-temperature properties of a hydrogen monolayer on
graphene (graphane). Also, the dynamics of free-standing graphene
multilayers may display interesting quantum features at low temperature.

\begin{acknowledgments}
The authors acknowledge the support of J. H. Los in the implementation 
of the LCBOPII potential, and
E. Chac\'on is thanked for inspiring discussions.
This work was supported by Direcci\'on General de Investigaci\'on,
MINECO (Spain) through Grants FIS2012-31713 and FIS2015-64222-C2
\end{acknowledgments}

\end{document}